\documentclass[aps,prd,reprint,superscriptaddress,nofootinbib,nobibnotes,longbibliography]{revtex4-2}

\newcommand{\RomanNumeralCaps}[1]
    {\MakeUppercase{\romannumeral #1}}

\usepackage{xspace}
\usepackage{amssymb}
\usepackage{amsmath}
\usepackage[utf8]{inputenc}
\usepackage{graphicx}
\usepackage{cancel}
\usepackage{bbm}
\usepackage{color}
\usepackage{lineno}
\usepackage{dcolumn}   
\usepackage{bm}        
\usepackage{amssymb}   
\usepackage{amsmath,amssymb}
\usepackage{comment}
\usepackage[dvipsnames]{xcolor}
\usepackage[colorlinks = true,
            linkcolor  = RoyalBlue,
            urlcolor   = RoyalBlue,
            citecolor  = RoyalBlue]{hyperref}
            
\usepackage[utf8]{inputenc}
\usepackage{amsmath}
\usepackage{graphicx}
\usepackage{color}
\usepackage{graphicx}
\newcommand{\pt}           {\ensuremath{p_{\rm T}}\xspace}

\newcommand{\twoHnn}       {$\sqrt{s_{\mathrm{NN}}}=200$~GeV\xspace}
\newcommand{\twosevensixnn}{$\sqrt{s_{\mathrm{NN}}}=2.76$~TeV\xspace}

\newcommand{\fivefourfournn}{$\sqrt{s_{\mathrm{NN}}}=5.44$~TeV\xspace}

\newcommand{\fluidum}      {\textsc{Fluid{\it u}M~}}

\newcommand{\trento}       {\textsc{TrENTo~}}
\newcommand{\fastreso}     {\textsc{FastReso~}}

\begin{document}
\title{Quantification of the low-$p_{\rm T}$ pion excess in heavy-ion collisions \\at the LHC and top RHIC energy}
\date{\today}

\author{P.~Lu}
\email[]{pengzhong.lu@cern.ch }
\affiliation{GSI Helmholtzzentrum f{\"u}r Schwerionenforschung, 64291 Darmstadt, Germany}
\affiliation{University of Science and Technology of China, China}

\author{R.~Kavak}
\email[]{kavak@physi.uni-heidelberg.de }
\affiliation{Physikalisches Institut, Universit{\"a}t Heidelberg, 69120 Heidelberg, Germany}

\author{A.~Dubla}
\email[]{a.dubla@gsi.de}
\affiliation{GSI Helmholtzzentrum f{\"u}r Schwerionenforschung, 64291 Darmstadt, Germany}

\author{S.~Masciocchi}
\email[]{s.masciocchi@gsi.de}
\affiliation{Physikalisches Institut, Universit{\"a}t Heidelberg, 69120 Heidelberg, Germany}
\affiliation{GSI Helmholtzzentrum f{\"u}r Schwerionenforschung, 64291 Darmstadt, Germany}

\author{I.~Selyuzhenkov} 
\email[]{ilya.selyuzhenkov@gmail.com}
\affiliation{GSI Helmholtzzentrum f{\"u}r Schwerionenforschung, 64291 Darmstadt, Germany}

\begin{abstract}

While the abundances of the final state hadrons in relativistic heavy-ion collisions are rather well described by the thermal particle production, the shape of the transverse momentum, \pt, distribution below $\pt\approx500$~MeV$/c$, is still poorly understood.
We propose a procedure to quantify the model-to-data differences using Bayesian inference techniques, which allows for consistent treatment of the experimental uncertainties and tests the completeness of the available hydrodynamic frameworks.
Using relativistic fluid framework \fluidum with PCE coupled to \trento initial state and \fastreso decays, we analyse \pt distribution of identified charged hadrons measured in heavy-ion collisions at top RHIC and the LHC energies, and identify an excess of pions produced below $\pt\approx500$~MeV$/c$.
Our results provide new input for the interpretation of the pion excess as either missing components in the thermal particle yield description or as an evidence for a different particle production mechanism.

\end{abstract}

\maketitle

\section{Introduction}

Experiments involving heavy-ion collisions at ultra-relativistic energies, conducted at the Relativistic Heavy Ion Collider (RHIC) and the Large Hadron Collider (LHC), aim at studying a new state of matter known as the quark--gluon plasma (QGP)~\cite{Busza:2018rrf, ALICE:2022wpn, STAR:2005gfr, PHENIX:2004vcz}.
Viscous hydrodynamics is remarkably successful at describing a wide range of observables and has become the “standard model” for the evolution of ultra-relativistic heavy-ion collisions~\cite{Shen:2020mgh,Schenke:2020mbo,Nijs:2020roc}.
Abundances of the final state hadrons contain important information about the dynamics of the QGP created in relativistic heavy-ion collisions.
While the measured integrated yields are rather well understood within the picture of the thermal particle production, the shape of the transverse momentum distribution, in particular at low transverse momentum (\pt) below 500~MeV$/c$, is still poorly understood within the state-of-the-art hydrodynamic model calculations~\cite{Nijs:2020roc,Schenke:2020mbo,Vermunt:2023zsk,Heffernan:2023utr}.
The particle production at low transverse momentum is associated with the long-distance scales, which are accessible in heavy-ion collisions and out of reach in hadronic interactions.
Its enhancement could indicate transverse momentum and particle species yield redistribution of the thermally produced hadrons due to conventional phenomena~\cite{Schnedermann:1993ws,Huovinen:2016xxq,Andronic:2017pug,Lo:2017sux,Andronic:2018qqt,Mazeliauskas:2018irt}, which is not yet implemented in the current state-of-the-art hydrodynamic models, or new physics phenomena~\cite{Busza:1992sr,Shuryak:1990ie,Kataja:1990tp,Gavin:1991ki,Lee:1990sk,Simon-Gillo:1994ghv,Sollfrank:1991xm,Brown:1991en,Blaizot:1996js,Bjorken:1997re,Petersen:1999jc,Begun:2015ifa,Begun:2016cva,Grossi:2021gqi}.

An enhancement at low transverse momentum was first observed at the ISR in high multiplicity pp and $\alpha$--$\alpha$ collisions when compared to minimum bias pp collisions~\cite{CERN-Heidelberg-Lund:1984uxp} and in p--A collisions at Fermilab and CERN~\cite{Chaney:1979bz,Garbutt:1977zu}, and later in A--A collisions at the AGS and CERN~\cite{NA35:1988rpj,Simon-Gillo:1994ghv,E-810:1991sbm,Hemmick:1994rn,Gonin:1994wgt}. 
A less than 10\% enhancement at low-\pt was observed for midrapidity pions in both p--Pb and A--A collisions by the NA44 Collaboration~\cite{NA44:1992tww}.
The low-\pt enhancement in A--A collisions showed no target size dependence and was smaller for pions at midrapidity compared to target rapidity~\cite{Simon-Gillo:1994ghv}.
Intriguing explanations proposed at that time included exotic behavior in dense hadronic matter~\cite{Shuryak:1990ie,Kataja:1990tp,Gavin:1991ki}, the decay of quark matter droplets~\cite{VanHove:1988qt}, collective effects~\cite{Lee:1990sk}, baryonic and mesonic resonance decays~\cite{Simon-Gillo:1994ghv,Sollfrank:1991xm,Brown:1991en}, and 
the possible formation of a transient state with partially restored chiral symmetry in the early stage of the heavy-ion collision~\cite{Blaizot:1996js,Bjorken:1997re,Petersen:1999jc}.

The mechanism of low-\pt pion production remains an open question for experiments at the modern heavy-ion colliders.
An excess is seen when comparing low-\pt pion yield measured by the ALICE~\cite{ALICE:2013mez,ALICE:2019hno,ALICE:2021lsv,ALICE:2016dei} at the LHC and the PHENIX and STAR~\cite{PHENIX:2003iij,PHENIX:2001vgc,STAR:2003jwm,STAR:2008med} at RHIC to the hydrodynamic model calculations~\cite{Kolb:2003dz,Vermunt:2023zsk,Devetak:2019lsk,Nijs:2020roc,Dubla:2018czx,Gale:2012rq,McDonald:2016vlt,Mazeliauskas:2019ifr,Heffernan:2023utr,Heffernan:2023gye}.
An indication of the excessive pion yield, though with large uncertainties, is also visible from the comparison of the thermal model fits to measured integrated yields of different particle species~\cite{Andronic:2017pug}.
While in most experiments at RHIC and the LHC the pion \pt spectra are measured only above \pt~=~0.1-0.2~GeV$/c$, the PHOBOS experiment at RHIC~\cite{ByPHOBOS:2004dqp,PHOBOS:2006zpw} measured it down to the \pt=30--50~MeV$/c$.
An extrapolation of the blast-wave model~\cite{PHOBOS:2006zpw}, fitted to PHOBOS experimental data in the intermediate \pt region, revealed no significant increase in kaon and proton production when compared to low-\pt data.
However, the same extrapolation showed a possible enhancement in pion production at very low \pt.

The low-\pt pion excess may arise from physics mechanisms not accounted for in the current hydrodynamic model simulations, like Bose--Einstein condensation~\cite{Begun:2015ifa, Begun:2016cva}, increased population of resonances~\cite{Schnedermann:1993ws}, treatment of the finite width of $\rho$ meson~\cite{Huovinen:2016xxq}, or critical chiral fluctuations~\cite{Grossi:2021gqi}.
Quantification of the low-\pt pion yield excess is important for both, the improvement of fluid dynamic modeling and the search for new particle production mechanisms in heavy-ion collisions.
On the experimental side, the proposed next-generation detector ALICE 3 at the LHC~\cite{ALICE:2022wwr}, which combines excellent particle identification capabilities, a unique pointing resolution, and large rapidity coverage, will allow measurements below \pt~$\sim$~100~MeV$/c$.

To advance in understanding of the mechanism for low \pt particle production we propose a procedure to systematically quantify the model-to-data differences using modern Bayesian inference analysis techniques, which allows for consistent treatment of the experimental uncertainties.
In this paper, we deploy a procedure based on the relativistic fluid framework \fluidum with partial chemical equilibrium (PCE) coupled to \trento initial state and \fastreso decays to analyse \pt distributions of charged pions, kaons and protons measured in collisions of Pb--Pb at \twosevensixnn~\cite{ALICE:2013mez}, Xe--Xe at \fivefourfournn~\cite{ALICE:2021lsv}, and Au--Au at \twoHnn~\cite{PHENIX:2003iij,STAR:2008med}.
Our results demonstrate the power of the proposed procedure to exploit the precision of the current experimental data in the search for limitations and improvements in the available state-of-the-art hydrodynamic model frameworks.

\section{Modelling of heavy-ion collisions}
\label{sec:setup}
Our model for simulating high-energy nuclear collisions combines three distinct components.
The \trento model~\cite{Moreland:2014oya} was utilized for the initial conditions, while the \fluidum model with a PCE implementation~\cite{Floerchinger:2018pje}, featuring a mode splitting technique for fast computations, was used for the relativistic fluid dynamic expansion with viscosity. 
Additionally, the \fastreso code~\cite{Mazeliauskas:2018irt} was used to take resonance decays into account. 

The \trento model involves positioning nucleons with a Gaussian width $w$ using a fluctuating Glauber model, while ensuring a minimum distance $d$ between them. 
Each nucleon contains $m$ randomly placed constituents with a Gaussian width of $v$. 
\trento uses an entropy deposition parameter $p$ that interpolates among qualitatively different physical mechanisms for entropy production~\cite{Moreland:2014oya}. Furthermore, additional multiplicity fluctuations are introduced by multiplying the density of each nucleon by random weights sampled from a gamma distribution with unit mean and shape parameter $k$.
For this study, the \trento parameters are set based on \cite{Moreland:2018gsh}.
The inelastic nucleon--nucleon cross sections are taken from the measurements by the ALICE and PHENIX Collaborations~\cite{ALICE-PUBLIC-2018-011,PHENIX:2015tbb}. The Pb and Au ions are sampled from a spherically symmetric Woods--Saxon distribution, while the Xe ion comes from a spheroidal Woods--Saxon distribution with deformation parameters $\beta_2=0.21$ and $\beta_4=0.0$~\cite{Bally:2021qys}. 

The software package \fluidum~\cite{Floerchinger:2018pje}, which utilizes a theoretical framework based on relativistic fluid dynamics with mode expansion~\cite{Floerchinger:2013rya, Floerchinger:2013hza, Floerchinger:2014fta}, is used to solve the equations of motion for relativistic fluids. The causal equations of motion are obtained from second-order Israel--Stewart hydrodynamics~\cite{Floerchinger:2017cii}.
As in our previous work \cite{Vermunt:2023zsk}, we are interested in examining the azimuthally averaged transverse momentum spectra of identified particles at midrapidity. Therefore, we do not consider azimuthal and rapidity-dependent perturbations and only require the background solution to the fluid evolution equations, neglecting terms of quadratic or higher order in perturbation amplitudes. 

The Cooper--Frye procedure is used to convert fluid fields to the spectrum of hadron species on a freeze-out surface, which in our work is assumed to be a surface of constant temperature~\cite{Cooper:1974mv}. 
As in our previous work~\cite{Vermunt:2023zsk}, the
hadronic phase, after the chemical freeze-out and before the kinetic freeze-out, i.e.~$T_{\rm kin} < T < T_{\rm chem}$, is modeled by a concept of partial chemical equilibrium (PCE), which replaces the need for a hadronic after-burner in the simulation. Our description follows the work described in Refs.~\cite{Bebie:1991ij, Huovinen:2007xh, Motornenko:2019jha}, in which different particle species in a hadronic gas are treated as being in chemical equilibrium with each other, while the overall gas is not.
During the PCE, the mean free time for elastic collisions is still smaller than the characteristic expansion time of the expanding fireball, thereby keeping the gas in a state of local kinetic equilibrium. The chemical equilibrium is not maintained if the mean free path of the inelastic collisions exceeds this threshold. 
On the kinetic freeze-out surface, we take the particle distribution function to be given by the equilibrium Bose--Einstein or Fermi--Dirac distribution (depending on the species), modified by additional corrections due to bulk and shear viscous dissipation~\cite{Teaney:2003kp,Paquet:2015lta} and decays of unstable resonances~\cite{Mazeliauskas:2018irt}. 
We use a list of approximately 700 resonances from Refs.~\cite{Alba:2017mqu, Alba:2017hhe, Parotto_private}.

As described in Ref.~\cite{Vermunt:2023zsk}, our central framework revolves around certain free parameters: the overall normalization constant ${\rm Norm}$, $(\eta/s)_{\rm min}$
and $\left(\zeta/ s\right)_{\rm max}$ in the shear and bulk viscosity to entropy ratio parametrizations, the initial fluid time $\tau_0$, and the two freeze-out temperatures $T_{\rm kin}$ and $T_{\rm chem}$.
With our Bayesian inference analysis, we simultaneously determine these six model
parameters within predefined intervals (refer to Tab.~\ref{tab:predefinedParameters}). 
These intervals are based on physical considerations and knowledge from previous studies~\cite{Devetak:2019lsk, Vermunt:2023zsk, Ryu:2017qzn, Andronic:2017pug, ALICE:2013mez}.
It is worth mentioning that we have confirmed \textit{a posteriori} that the optimal values fall within these intervals rather than on their boundaries, and in cases where no clear convergence was obtained, larger intervals were employed.
Although \fluidum is recognized for its fast execution speeds, the extensive parameter exploration involved in Bayesian analyses necessitates an approach to speed up the simulations. Our approach is based on the usage of an ensemble of artificial neural networks (ANNs) to emulate our model calculations. The training necessitates large datasets
to achieve the required accuracy for replacing the simulation outputs. For each collision system, we use the outputs of
ten thousand complete model calculations, with parameters distributed within the ranges presented in Tab.~\ref{tab:predefinedParameters}. The parameter values are generated using Latin hypercube sampling, which ensures a uniform density. With this large population of initial points in the parameter space, the emulator uncertainties result in a few percent. We refer readers to Refs.~\cite{Vermunt:2023zsk,Seemann2022}, which provide extensive discussions. The posterior density is inferred from a probabilistic model, we use the numerical Markov-Chain Monte Carlo (MCMC) method~\cite{Foreman-Mackey:2012any}, which is an efficient approach for exploring the probability space.
Without clear guidance on how to precisely handle the degree of correlation in the experimental systematic uncertainties, the Bayesian inference analysis is performed assuming the experimental systematic uncertainties uncorrelated among the different particle species and transverse momentum intervals. 

\begin{table}
\centering
\caption{Ranges for the model parameters across three collision systems. The normalization constant and the initial fluid time are treated as system-dependent parameters.}
\label{tab:predefinedParameters}
\begin{tabular}{c|cccc}
\toprule
 & Pb--Pb  &  Xe--Xe & Au--Au \\ 
\hline
$(\zeta / s)_{\rm max}$ & \multicolumn{3}{c}{$10^{-4}-0.3$}  \\ 
$(\eta / s)_{\rm min}$ & \multicolumn{3}{c}{$0.08-0.78$} \\ 
$T_{\rm chem}$ (MeV) & \multicolumn{3}{c}{$130-155$} \\ 
$T_{\rm kin}$ (MeV)& \multicolumn{3}{c}{$110-140$} \\ 
Norm & $20-80$ &  $50-150$& $3-80$ \\ 
$\tau_0$ (fm$/c$)& $0.1-3.0$ &  $0.5-7.0$ & $0.5-3.0$\\ 
\toprule
\end{tabular}
\end{table}

\section{Determination of the optimal fitting range}
\label{sec:optimalPtRanges}

The pion, kaon, and proton \pt spectra across various collision centrality classes measured by the ALICE Collaboration at the LHC and by the PHENIX and STAR Collaborations at RHIC in different colliding systems and center-of-mass energies, namely Pb--Pb collisions at \twosevensixnn~\cite{ALICE:2013mez}, Xe--Xe collisions at \fivefourfournn~\cite{ALICE:2021lsv}, and Au--Au collisions at \twoHnn~\cite{PHENIX:2003iij,STAR:2008med} are used in this work. For the RHIC energy, the $\overline{\rm p}$ spectra are used in the Bayesian inference analysis because in our model the stopping of the baryons from the colliding nuclei (baryon transport at midrapidity) is not included.
Differently from our previous work~\cite{Vermunt:2023zsk}, we expanded the Bayesian inference analysis to include multiple centrality intervals covering the range 0--40\% for all collision systems.
We highlight that our objective in this work is to systematically quantify the low-\pt pion excess and examine its possible dependence on collision centrality, collision energy, and colliding nuclei, rather than constraining physical parameters of the QGP across these three collision systems. For this reason, we have run the full framework separately for each centrality interval and collision system, without attempting to perform a global fit using all available data. 

To determine the optimal $p_{\rm T}^{\pi}$ range for fitting experimental measurements, and consequently to compute the low-\pt pion excess, we performed a Bayesian inference analysis varying each time the \pt interval of the pion spectra, while keeping them for kaon and proton spectra fixed ($\pt^{\rm K, p} <$ 2.0 GeV$/c$). Because of the large difference in masses, simultaneous inference of pion, kaon, and proton spectra was employed to achieve convergence of the model parameters.
Initially, we optimized the starting $p_{\rm T}^{\pi}$ within the range $x_1 < p_{\rm T}^{\pi} <$ 2.0 GeV$/c$, where $x_1$ ranged from 0.1 to 1.0 GeV$/c$. This range was chosen to ensure an adequate number of \pt intervals for the Bayesian inference procedure, with the upper limit for $x_1$ set at 1.0 GeV$/c$. Subsequently, we optimized the ending \pt by fitting within 0.5 $< p_{\rm T}^{\pi} < x_2$ GeV$/c$, where $x_2$ ranged from 2.0 to 3.0 GeV$/c$.

\begin{figure*}[tb!]
    \centering
    \includegraphics[width=1\linewidth]{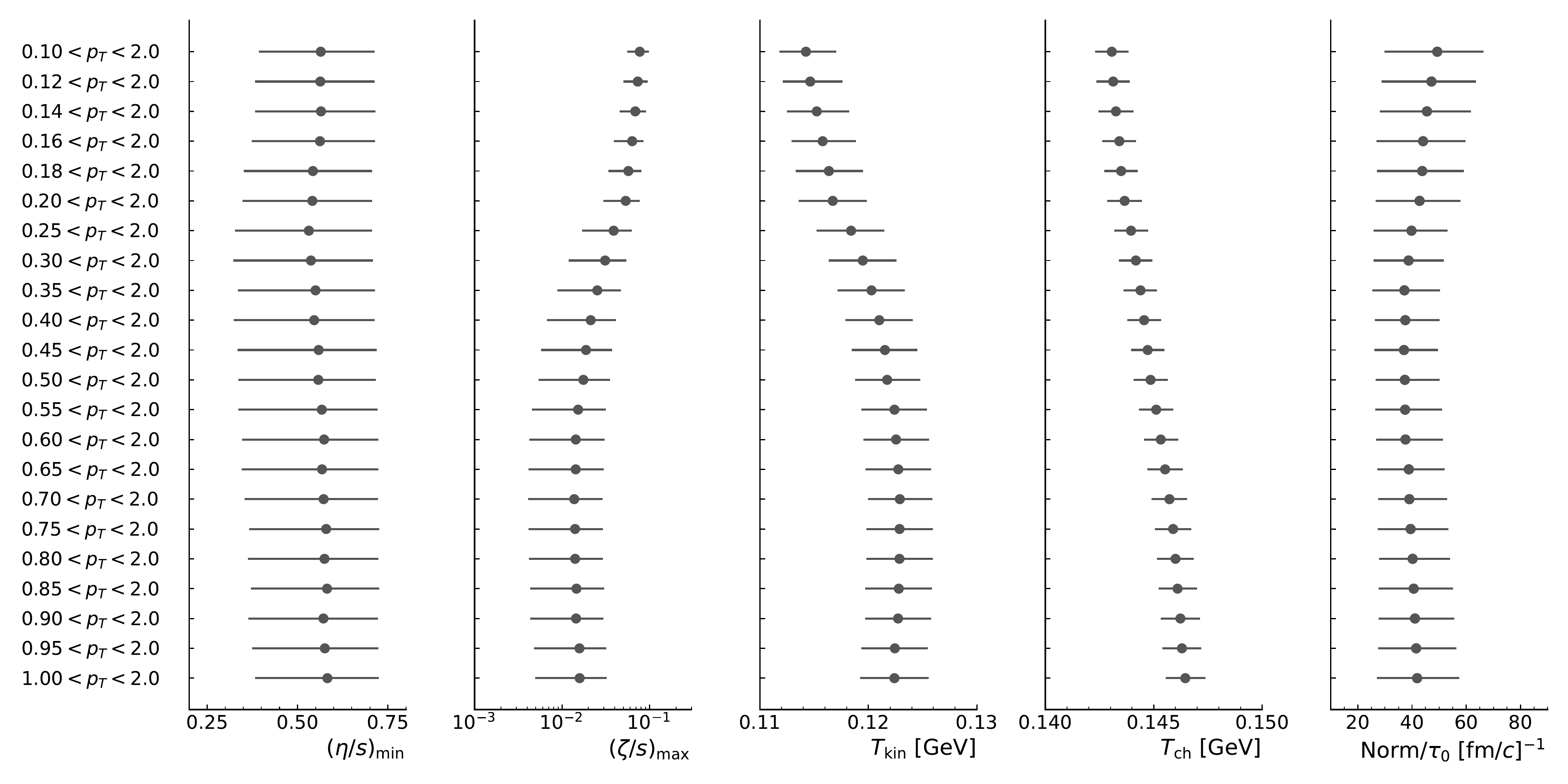}
    \includegraphics[width=1\linewidth]{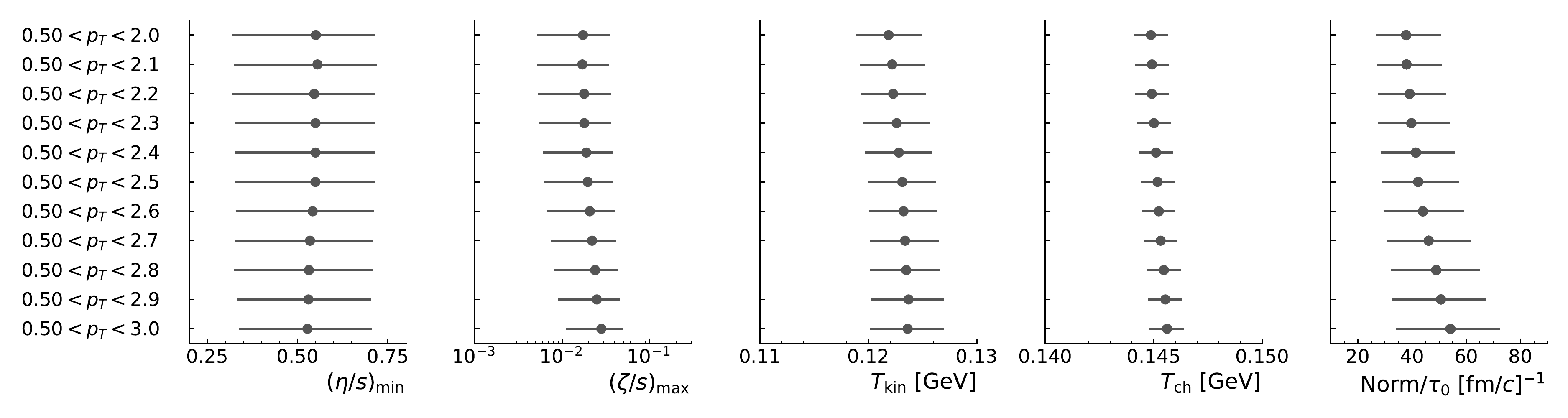}
    \caption{
      Parameter values within different \pt fitting ranges for the $0$--$5\%$ centrality class in Pb--Pb collisions at \twosevensixnn. The upper panel displays the variation of the starting $p_{\rm T}^{\pi}$ from 0.1 GeV$/c$ to 1.0 GeV$/c$. Conversely, the lower panel illustrates the variation of the ending $p_{\rm T}^{\pi}$ from 2.0 GeV$/c$ to 3.0 GeV$/c$. The error bars in the figure denote 68\% confidence intervals of the marginalized Bayesian posterior distributions for each model parameter.}
    \label{ptoptimisation}
\end{figure*}

Although the constraint of the QGP physical parameters is not the main focus of this research, it is crucial to monitor their performance and convergence while optimizing the pion \pt range. This ensures that the chosen \pt range in the Bayesian inference procedure leads to convergence.
In Fig.~\ref{ptoptimisation} the six key parameters for the $0$--$5\%$ centrality class in Pb--Pb collisions at \twosevensixnn are shown. The Norm and \(\tau_0\) parameters are depicted in a ratio format (\(\text{Norm}/\tau_0\)) because in our model the expected entropy density profile is obtained using their ratio~\cite{Devetak:2019lsk, Vermunt:2023zsk}. 
The top panel corresponds to the starting $p_{\rm T}^{\pi}$ optimization procedure, while the bottom panel focuses on ending $p_{\rm T}^{\pi}$ optimization. The values reported represent the median of the marginalized Bayesian posterior distributions for each model parameter, while error bars denote the $68$\% confidence interval.

All parameters exhibit varying stability across the fitting ranges. In the top panel, the parameters converge to stable values when $x_1$ exceeds the \pt threshold of approximately 0.5 GeV$/c$. This indicates that using the low-\pt pion region ($x_1 <$ 0.5 GeV$/c$) in the Bayesian inference procedure would introduce instabilities in constraining the physical parameters, demonstrating that a fluid dynamic framework cannot capture the experimentally measured low-\pt pion spectra.
In the lower panel of Fig.~\ref{ptoptimisation}, the parameters start deviating from the converged values again when the Bayesian procedure includes the spectra values for $\pt^{\pi} >$ 2.0 GeV$/c$. 
As we move to higher \pt, it is anticipated that particles are no longer predominantly produced thermally.
We attribute this to the limit of the applicability of the fluid dynamics description at high \pt (emerging contribution from hard processes) and concluded that our results for the $\pt<2.5$~GeV$/c$ are stable and not subject to overfitting or overtraining.
Instead, contributions from hard partonic scattering processes become more pronounced, and the effects of partonic energy loss begin to dominate the spectral shape.
On top of the parameter instabilities, it was observed that even when either the low-\pt or high-\pt spectra are included in the Bayesian procedure, the \fluidum calculations fail to replicate the experimental data accurately. This results in significant discrepancies between the data and the model observed both at low and high \pt.
The same study was also conducted for the $30$--$40\%$ centrality interval in Pb--Pb collisions at \twosevensixnn to verify the consistency of the findings. Comparable performances were observed across the different centrality intervals analyzed. As a result, the optimal \pt range with respect to a fluid dynamic description was established to be $0.5 < p_{\rm T}^{\pi} <$ 2.0 GeV$/c$ for all centrality intervals and collision systems.

\begin{figure*}[th!]
  \centering
  \includegraphics[width=1\linewidth]{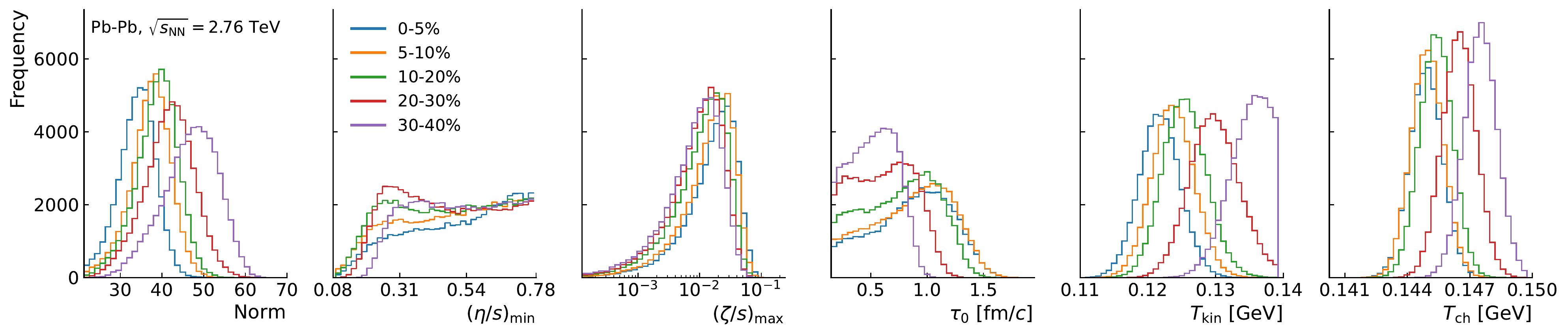}
  \caption{Marginal posterior distributions of the model input parameters for the five analyzed centrality classes in Pb--Pb collisions at \twosevensixnn.}
  \label{posteriros}
\end{figure*}

In Fig.~\ref{posteriros} the Bayesian posterior distributions of the model input parameters utilized in this analysis for all centrality classes in Pb--Pb collisions at \twosevensixnn are reported. 
It is important to note that we have confirmed the posterior distributions fall within the prior interval specified in Table \ref{tab:predefinedParameters}, rather than on its boundaries.
This marginal distribution plot illustrates that the parameters exhibit a high degree of consistency across the various centrality classes, lying within one standard deviation.
Notably, the parameter $T_{\rm kin}$ shows a systematic shift in its median value towards more peripheral collisions.
This observation aligns with previous findings obtained with the usage of a Blastwave fit~\cite{ALICE:2022wpn}, and it can be interpreted as a possible indication of a more rapid expansion towards central collisions and with the expectation of a shorter-lived fireball with stronger radial pressure gradients in more peripheral collisions.
As discussed in \cite{Vermunt:2023zsk}, the $(\eta / s)_{\rm min}$ remains unconstrained, which we attributed to the limited sensitivity of the current observables to the shear viscosity of the system.

This study determined that the optimal \pt range for the pion \pt spectra is $0.5 < p_{\rm T}^{\pi} <$ 2.0 GeV$/c$. This range is recommended when similar Bayesian analyses are performed to constrain physical QGP parameters and use \pt differential pion variables.
Having established this \pt interval, the one required to quantify the low-\pt pion excess is consequently defined as $0.1 < p_{\rm T}^{\pi} <$ 0.5 GeV$/c$ for the LHC and $0.2 < p_{\rm T}^{\pi} <$ 0.5 GeV$/c$ for RHIC.

\section{Low-\pt pion yields }

\begin{figure*}[ht!]
  \centering
  \includegraphics[width=.95\linewidth]{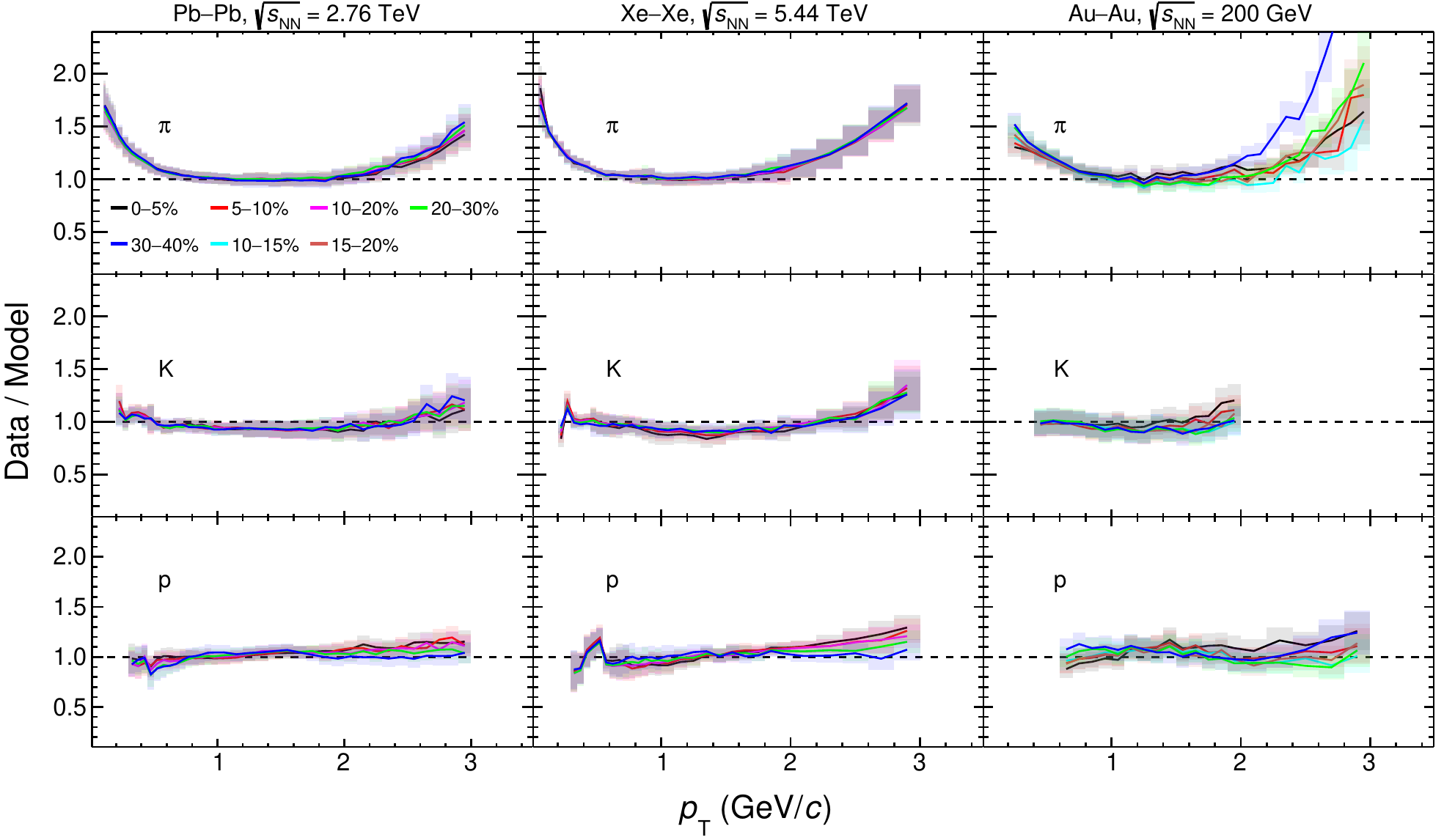}
  \caption{
  Differential yields of pions ($\pi$), kaons (K), and protons (p) over model spectra for 0--40\% centrality classes in Pb--Pb collisions at \twosevensixnn~\cite{ALICE:2013mez},
  Xe--Xe collisions at \fivefourfournn~\cite{ALICE:2021lsv}, and Au--Au collisions at\twoHnn~\cite{PHENIX:2003iij}, respectively. From top to bottom, each row corresponds to particle type, and from left to right, each column showcases one of the three collision systems under consideration. Within every panel, ratios are segmented into five (six) centrality classes ranging from 0\% to 40\%. The bands represent the statistical and systematic uncertainties of the data, summed in quadrature.}
  \label{dataOverModelRatios}
\end{figure*}

In Fig.~\ref{dataOverModelRatios} the ratios of the experimental spectra over model calculations for pions, kaons, and protons are shown. 
For those ratios, the model calculations are performed using the Maximum a Posteriori (MAP) estimates of the parameters~\cite{Moreland:2014oya}.
The MAP estimate refers to the set of model parameters corresponding to the mode of the posterior distribution, representing the point in parameter space with the highest posterior probability. Given that we use uniform priors in our Bayesian inference, the MAP values are equivalent to those that maximize the likelihood function.
The ratios are arranged in rows per particle type and columns per collision system. The bands depict experimental statistical and systematic uncertainties combined in quadrature. \fluidum calculations yield nearly flat data-over-model ratios compatible with unity within one standard deviation across the entire \pt spectrum for pion, kaons, and protons for all centrality intervals and collisions systems in the intervals used in the Bayesian analysis.

We recall that for kaons and protons the \pt interval used for the Bayesian analysis is \pt $<$ 2.0 GeV$/c$, while for pions it is 0.5 $<$ \pt $<$ 2.0 GeV$/c$. 
For $\pt >$ 2.0 GeV$/c$, the model calculations for all hadrons start to deviate from experimental measurements, suggesting that the higher \pt domain may not be predominantly governed by soft processes, which can typically be described by fluid dynamic calculations.
The observed deviations are larger for pions with respect to heavier particles, supporting the idea that hadrons originate from a fluid with a unified velocity field.
 
In the low-\pt range ($\pt <$ 0.5 GeV$/c$), unlike kaons and protons, the data-over-model ratios for pions exceed unity across all centrality classes and collision systems, indicating a systematic pion production excess in the experimental measurements with respect to the fluid dynamic production. As discussed in the previous section, even when including the pion spectra in the Bayesian inference analysis, the fluid dynamic calculation is not able to capture this low-\pt interval.

\begin{figure}[th!]
  \centering
  \includegraphics[width=0.9\linewidth]{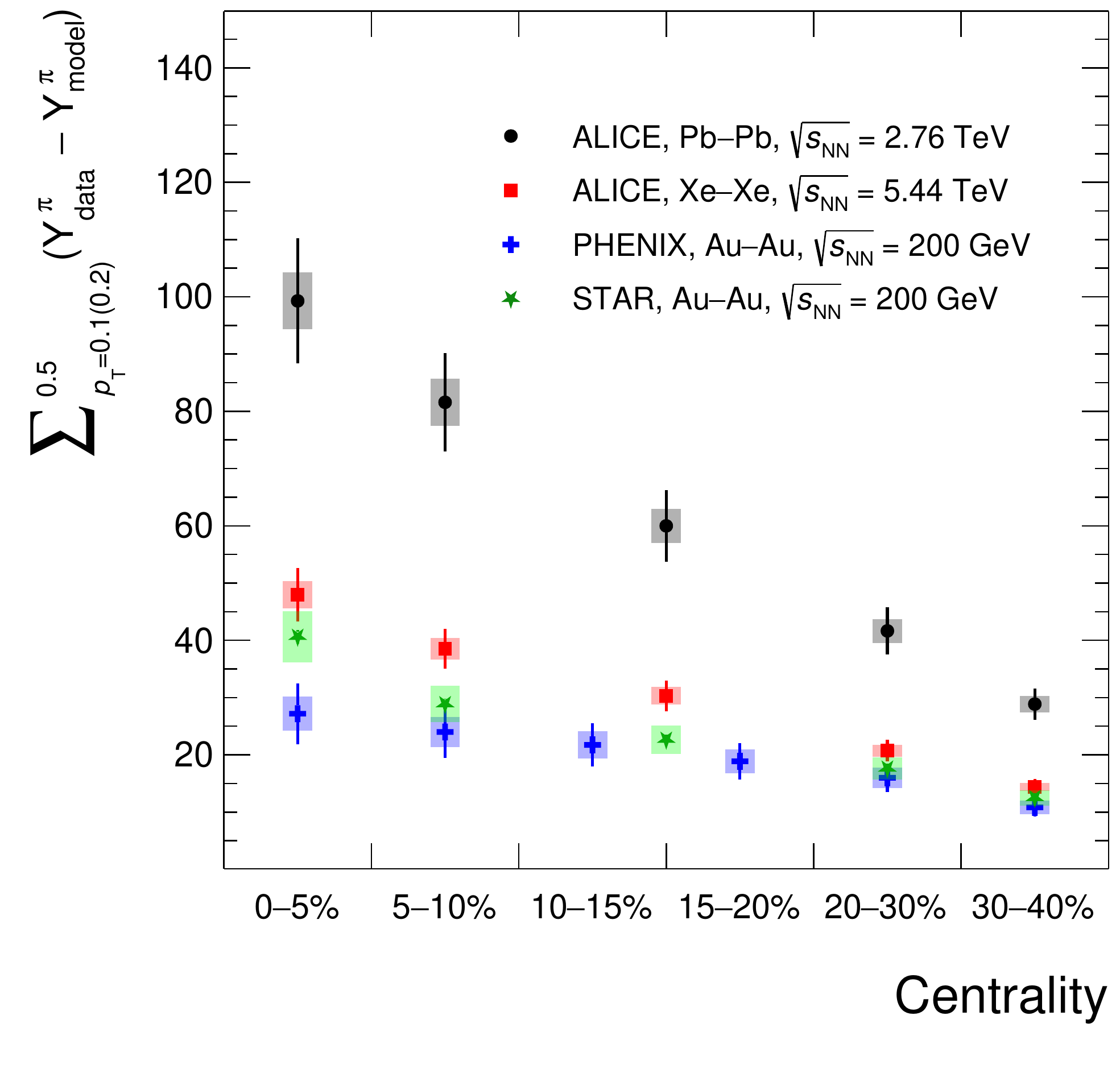}
  \caption{The integrated absolute single-charge pion excess in the range \(0.1(0.2) < p_{\text{T}} < 0.5\) GeV$/c$ as a function of centrality in different collision systems.
  The bars represent the experimental uncertainties and the bands represent the model uncertainties.}
  \label{excess}
\end{figure}

\begin{figure}[th!]
  \centering
  \includegraphics[width=0.9\linewidth]{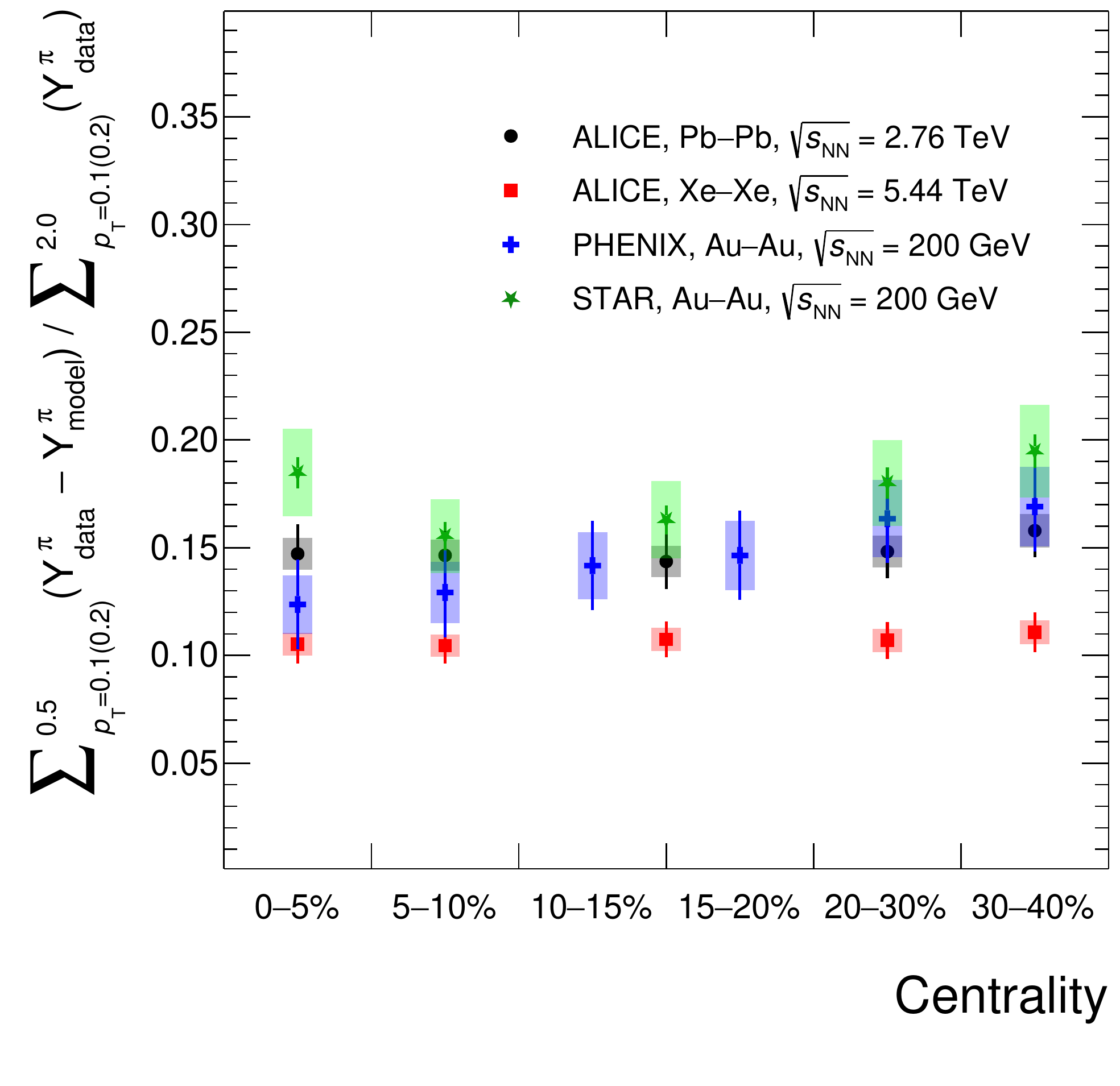}
  \caption{
  The excess of pions normalized to the integrated pion yields over \(0.1(0.2) < p_{\text{T}} < 2.0\) GeV/c for various centrality classes in the different collision systems. The bars represent the experimental uncertainties and the bands represent the model uncertainties.}
  \label{rel_excess}
\end{figure}

In Fig.~\ref{excess}, the pion excess, computed as the difference between the integral of the experimentally measured pion spectra in the interval $\pt <$ 0.5 GeV$/c$ and the integral of the pions computed within our framework in the same \pt interval, is shown for the three collision systems as a function of centrality.
It is important to notice that the excess is computed in two different \pt intervals for LHC and RHIC. At the LHC pion spectra are measured down to \pt = 0.1 GeV$/c$ while at RHIC down to \pt = 0.2 GeV$/c$.
This study focuses on the single-charge $\pi$ excess, utilizing $\pi^+$ \pt spectra in Pb--Pb at \twosevensixnn~\cite{ALICE:2013mez} and averaging $\pi^+$ and $\pi^-$ in Xe--Xe at \fivefourfournn~\cite{ALICE:2021lsv}.
As for the Au--Au system, the $\pi^-$ \pt spectra from PHENIX \cite{PHENIX:2003iij} are used to compute the excess.
For completeness, the pion excess is computed also utilizing the measured pion spectra from the STAR Collaboration (green markers) measured in Au--Au collisions at \twoHnn~\cite{STAR:2008med}.
Due to the limited \pt interval of the STAR measurement (0.2--0.75~GeV$/c$), it was not possible to perform an independent Bayesian analysis, and the pion excess is computed using the model calculation obtained by the inference analysis of the PHENIX data.
The excesses obtained using experimental data from the two Collaborations are compatible within the uncertainties.
The STAR measurements are available for the 10--20\% interval, while the PHENIX data for the centrality intervals 10--15\% and 15--20\%.
Despite different acceptance in rapidity of the STAR, PHENIX, and PHOBOS experiments at RHIC, all measurements are reported at midrapidity per unit of rapidity and no additional treatment is required when comparing our calculations to these data from different RHIC experiments. 
Therefore, to calculate the pion excess, the fluid calculations from the two finer centrality classes are averaged into a larger one.
The experimental uncertainties from measurements are reported as bars.
For consistency with the treatment of the systematic uncertainties in the fit, while computing the pion excess, the experimental systematic uncertainties are propagated as fully uncorrelated across \pt.
The total uncertainties represent the quadratic sum of experimental statistical and systematic uncertainties.
A decreasing trend in the excess from central to peripheral collisions is observed for all collision systems.
The significance of the excess is above 5 for all centrality classes and collision systems, specifically varying from 9.3 to 11.1 across centrality classes at the LHC energies.
We estimated the effect of treating the experimental systematic uncertainties as partially or fully correlated and the extracted pion excess remained compatible with our main result reported in the paper.
In the future, it will be beneficial to have experimental guidance on the degree of correlation of the uncertainties among \pt intervals and particle species of the experimental observables, which would enable a more thorough treatment of the systematic uncertainties in the analysis.

The uncertainties depicted by the bands in the figures originate from our model reflecting different sources of model and experimental uncertainty.
The uncertainties represent the spread in posterior distributions and the extrapolation in the parameter space performed by the neural network (NN) emulator. As illustrated in Fig.~\ref{dataOverModelRatios}, the MAP parameters provide a sufficient description of the data, indicating that the predominant source of model uncertainty arises from our NN emulator. Enhancements in the posterior distribution's precision could be achieved by conducting the calibration with an increased number of design points and a narrower range of parameter values to increase the density of the training points to reduce the interpolation uncertainty.

In Fig.~\ref{rel_excess}, the excess relative to the integral of the experimental data in the interval 0.1 $< \pt <$ 2.0 GeV$/c$ for the LHC and 0.2 $< \pt <$ 2.0 GeV$/c$ for RHIC is presented.
When calculating the relative excess the systematic uncertainties between the excess and the integrated pion yields are treated as correlated, which partially cancels them out.
For the STAR measurements, due to the limited \pt intervals, the PHENIX measurements are used as the denominator, hence no cancellation in the systematic uncertainties was possible. 
To obtain the yields for the 10--20\% centrality interval from PHENIX, the arithmetic average of the yields for 10--15\% and 15--20\% was used.
The relative excess remains constant as a function of centrality, with a consistent 10--20\% excess across different collision systems.
The computed relative excess indicates that fluid dynamic calculations account only for 80--90\% of the measured pion production in heavy-ion collisions. An excess yield is found in all collision systems and centrality ranges considered.

\begin{figure}[th!]
  \centering
  \includegraphics[width=0.95\linewidth]{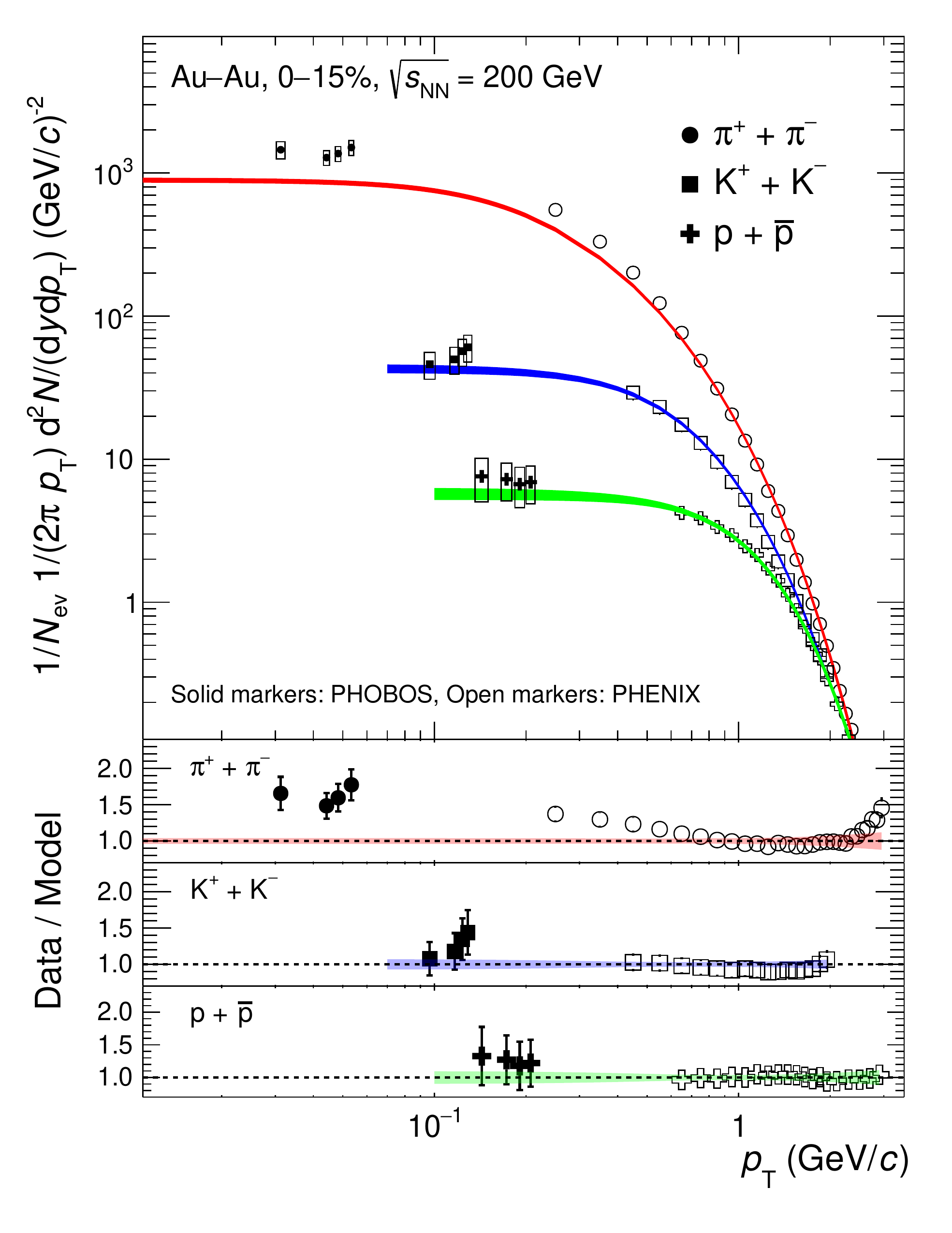}
  \caption{
  The invariant yields of pions, kaons, and protons as functions of \pt measured by the PHOBOS~\cite{PHOBOS:2006zpw} and the PHENIX~\cite{PHENIX:2003iij} Collaborations compared to the fluid dynamic calculations.}
  \label{LowPtRHIC}
\end{figure}

Having performed the Bayesian inference analysis utilizing the available RHIC data at \twoHnn, we can further compare our results with the measurements from the PHOBOS experiment at very low-\pt, to determine if a pion enhancement can also be quantified for \pt $<$ 100 MeV~\cite{ByPHOBOS:2004dqp}. 
The PHOBOS measurements are reported for the 0--15\% centrality interval close to midrapidity ($-0.1 < y < 0.4$). In Fig.~\ref{LowPtRHIC} the pion, kaon, and proton invariant yields are reported for the PHOBOS measurement at low-\pt (solid markers) and for the PHENIX data at larger \pt~\cite{PHENIX:2003iij} (open markers), in comparison with the MAP model calculation obtained via the Bayesian inference analysis previously described. To obtain the centrality interval 0--15\%, the results of our model and the PHENIX data for 0--5\%, 5--10\%, and 10--15\% were averaged. The experimental systematic uncertainties were propagated as correlated across the different centrality intervals.
To compute the sum of positively and negatively charged particles in our model, which predicts them in the same amount, the positively charged particle spectra were scaled by the experimentally measured numbers reported in Tab. \RomanNumeralCaps{9} of Ref.~\cite{PHENIX:2003iij}. This correction is significant only for the proton case due to the experimental difference in the measured proton and antiproton spectra.
In the bottom panels of Fig.~\ref{LowPtRHIC} the ratios to the various particle species are reported. No significant enhancement of kaons and protons is observed within the current experimental precision, while a deviation of about 50\% is observed for the low-\pt pion. 
This might indicate that the pion excess below \pt $<$ 0.1 GeV$/c$ saturates and does not keep rising to larger values.
However, to compute an integral of the full pion excess it is important to have experimental measurements covering the full \pt without having gaps within the measurement, which is envisioned by the proposed next-generation detector ALICE 3 at the LHC~\cite{ALICE:2022wwr}.

\section{Summary}

In summary, we propose a procedure to advance in understanding the mechanism for low \pt particle production in heavy-ion collisions, which allows to systematically quantify the model-to-data differences using modern Bayesian inference analysis techniques and consistently treat the experimental uncertainties.
We deploy this procedure using relativistic fluid framework \fluidum with PCE coupled to \trento initial state and \fastreso decays to analyse \pt distribution of charged pions, kaons, and protons measured in heavy-ion collisions at top RHIC and the LHC energies.
Despite the limited information about the correlation among systematic uncertainties in the data, our results indicate a systematic excess of pions produced below $\pt\approx500$~MeV$/c$ for both RHIC and LHC data.
Our results demonstrate the power of the proposed procedure to fully exploit the precision of the experimental data and search for limitations and improvements in the available state-of-the-art hydrodynamic model frameworks.
Further work, which is beyond the scope of this paper, is required for the interpretation of the observed low-\pt pion excess in terms of transverse momentum and particle species yield redistribution of the thermally produced particles, which are not yet modeled by the current state-of-the-art hydrodynamic models, or as evidence for a new particle production mechanism.

\textit{Acknowledgment}
This work is part of and supported by the DFG Collaborative Research Centre "SFB 1225 (ISOQUANT)".
Computational resources have been provided by the GSI Helmholtzzentrum f{\"u}r  Schwerionenforschung.

\bibliography{bibliography}

\end{document}